**Watching an uniformly moving source of light using a telescope and a frequency-meter**

**Bernhard Rothenstein and Ioan Damian**
"Politehnica" University of Timisoara, Dept. of Physics,
1, Piata Regina Maria, 1900 Timisoara, Romania

E-mail: bernhard_rothenstein@yahoo.com,
idamian@etv.utt.ro

*Abstract*
*We propose a scenario that involves a stationary observer who detects a point like source of light moving with constant velocity at a constant altitude, using a telescope and a frequency-meter. We derive a formula for the angular velocity at which we should rotate the axis of the telescope and a formula that relates the proper period at which the source emits successive wave crests and the proper period at which the stationary observer receives them.*

**1. Watching the moving source using a telescope**

The scenario we propose involves an observer R located at the origin O of the inertial reference frame K(XOY) who is handling a telescope. A point like source of light S' moves parallel to the OX axis with constant velocity $v$ ($\beta = \dfrac{v}{c}$). Let $h$ be the constant altitude (Figure 1). K' is the rest frame of the source in which we define its position by the space coordinates ($x'_0 = 0; y'_0 = h$). Source S' emits light continuously in all directions in space. At a time of emission $t_e$ we define the position of the source in K by the space coordinates

$x = c\beta t_e$                                                                 (1)
$y = h$ .                                                                (2)

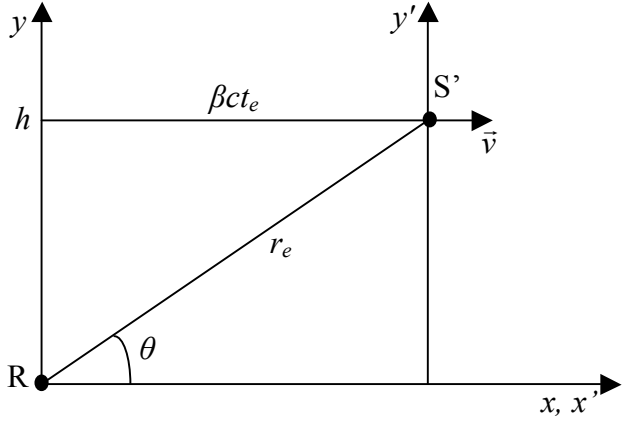

Fig. 1. Scenario for deriving the angular velocity at which we should rotate the axis of the telescope in order to detect a point like source of light that moves with constant velocity at a constant altitude.

The light signal emitted at $t_e$ arrives at the location of R when its clock reads $t_r$, related to $t_e$ by the obvious relation

$$t_r = t_e + \sqrt{\frac{h^2}{c^2} + \beta^2 t_e^2}. \qquad (3)$$

Solving for $t_e$ relation (3) leads to

$$t_e = \frac{t_r - \sqrt{\beta^2 t_r^2 + (1-\beta^2)\frac{h^2}{c^2}}}{1-\beta^2} \qquad (4)$$

where we have taken into account the solution with physical meaning ($t_r=0$; $t_e=-dc^{-1}$). Watching the moving source observer R should continuously rotate the axis of his telescope, which is at an angle $\theta$ relative to the positive direction of the OX axis at a time $t_r$ given by

$$\theta = arctg\frac{y}{x} = arctg\frac{\frac{h}{\beta c}(1-\beta^2)}{t_r - \sqrt{\beta^2 t_r^2 + (1-\beta^2)\frac{h^2}{c^2}}}. \qquad (5)$$

At a time $t_r$ the angular velocity at which R should rotate the axis of his telescope is

$$\omega = -\frac{h}{c}\left(\frac{1-\beta^2}{\beta}\right)\frac{A - \beta^2 t_r}{\left[(t_r - A)^2 + (\frac{h}{c})^2(\frac{1-\beta^2}{\beta})^2\right]A} \qquad (6)$$

where

$$A = \sqrt{\beta^2 t_r^2 + (1-\beta^2)\left(\frac{h}{c}\right)^2}. \qquad (7)$$

In Figure 2 we present the variation of the angular velocity $\omega$ with $t_r$ for a constant value of $\beta$ and different values of $\frac{h}{c}$, whereas in Figure 3 we present the variation of $\omega$ with $t_r$ for a constant value of $\frac{h}{c}$ and different values of $\beta$.

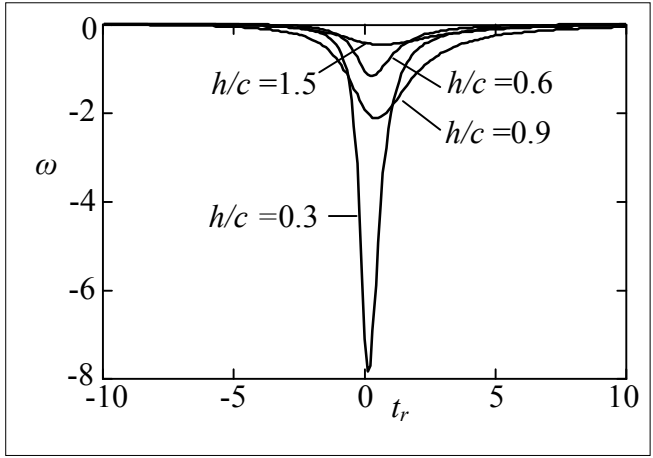

Fig. 2. The variation of the angular velocity ω with the reception time $t_r$ for the same value of $\beta = \dfrac{v}{c} = 0.6$ and different values of the altitude $\dfrac{h}{c}$.

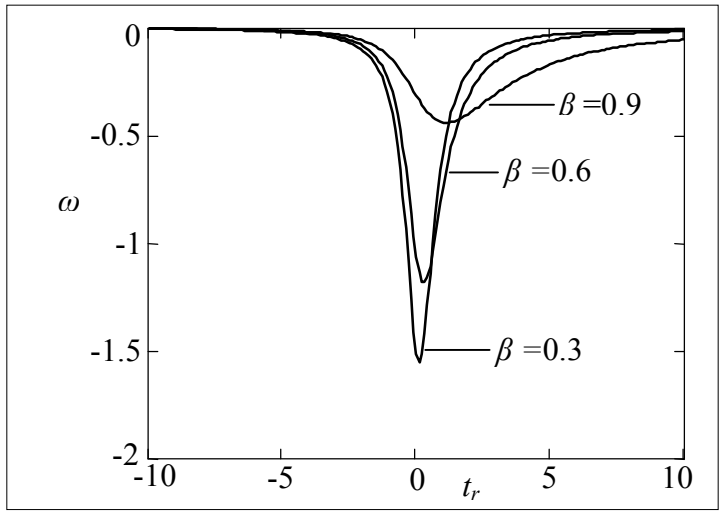

Fig. 3. The variation of the angular velocity ω with the reception time $t_r$ for a constant value of $\dfrac{h}{c} = 0.5$ and different values of $\beta = \dfrac{v}{c}$.

The derivation performed above does not invoke the special relativity theory but is in agreement with it by because distances measured perpendicular to the direction of relative motion are relativistic invariants as in Galileo's relativity.

**2. The time dependent Doppler formula**

The scenario presented in Figure1 leads to a Doppler formula as well. Presenting relation (3) as

$$t_r = t_e + \frac{r_e}{c} \tag{8}$$

where $r_e$ represents the magnitude of the position vector of S' in the K reference frame and differentiating both its sides we obtain

$$\frac{dt_r}{dt_e} = 1 + \beta \cos\theta_e . \tag{9}$$

In deriving relation (9) we have taken into account that, by definition,

$$\frac{dr_e}{dt_e} = v\cos\theta_e \tag{10}$$

represents the radial component of the instantaneous velocity of S'.

A clock commoving with the source measures a proper time interval $dt'_e$ related to $dt_e$ by the time dilation formula

$$dt_e = \frac{dt'_e}{\sqrt{1-\frac{v^2}{c^2}}} \qquad (11)$$

from which relation (10) becomes

$$\frac{dt_r}{dt'_e} = \frac{1+\beta\cos\theta_e}{\sqrt{1-\frac{v^2}{c^2}}}. \qquad (12)$$

If we consider that $dt_r$ represents the "very small" proper period at which R receives the successive light signals emitted by S' at a "very small" proper period of reception $dt'_e$ we can consider that relation (12) describes the Doppler Effect in the case of an oblique incidence of the successive light signals emitted by the source to the receiver. So far, we have considered that the periods $dt'_e$ and $dt_r$ are small enough that it can be considered that S' emits two successive light signals from the same point in space. Taking into account that

$$\cos\theta_e = \frac{\beta t_e}{\sqrt{\frac{h^2}{c^2}+\beta^2 t_e^2}} = \frac{\frac{\beta}{1-\beta^2}(t_r - A)}{\sqrt{\frac{h^2}{c^2}+\left(\frac{\beta}{1-\beta^2}\right)^2 (t_r - A)^2}} \qquad (13)$$

we can express relation (12) as a function of the reception time $t_r$.
In Figure 4 we present the variation of the Doppler factor

$$D = \frac{dt_r}{dt'_e} \qquad (14)$$

with $t_r$ for a constant value of $\beta$ and different values of $\frac{h}{c}$ whereas in Figure 5 we present its variation with $t_r$ for a constant value of $\frac{h}{c}$ and different values of $\beta$.

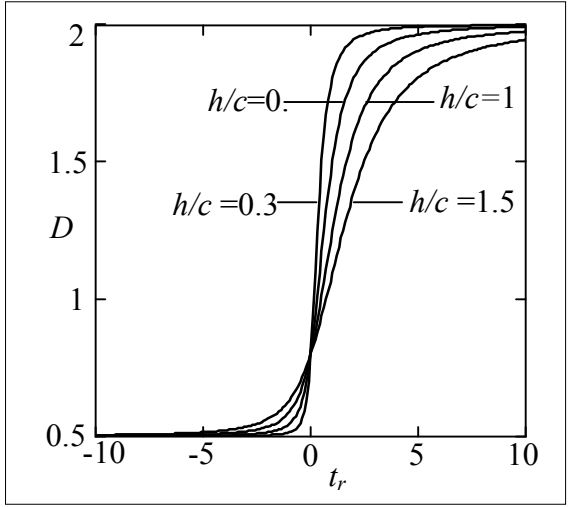

Fig. 4. The variation of the Doppler factor $D$ with $t_r$ for a constant value of β=0.6 and different values of $h/c$.

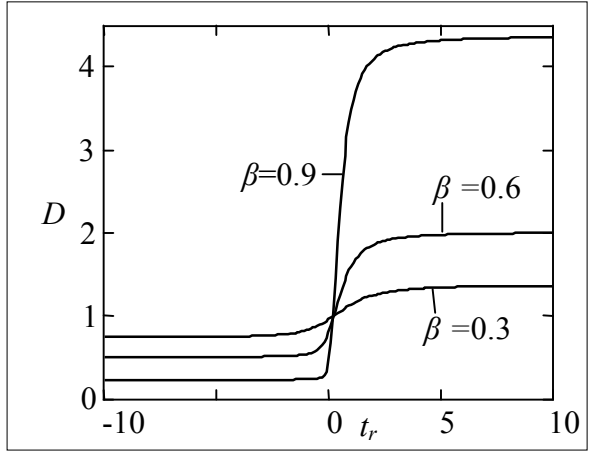

Fig. 5. The variation of the Doppler factor $D$ with $t_r$ for a constant value of h/c=0.5 and different values of β

We say that deriving the Doppler formula (14) we have made the "very small" period assumption. The plane wave assumption (very large source-receiver distance) leads to the same result[1] and so astronomers will consider that relation (14) and the results presented in Figure 4 and in Figure 5 hold exactly.

## 3. Conclusions

Watching a point like source that moves with constant velocity $\beta = \dfrac{v}{c}$ at a constant altitude $\dfrac{h}{c}$ using a telescope we should rotate it continuously. We show that the instantaneous angular velocity ω at which we should rotate the axis of the telescope depends on $v$ and on $\dfrac{h}{c}$ as well. Its peak value decreases if velocity and altitude increase. The derivation involves special relativity only by the invariance of the altitude, a distance measured perpendicular to the direction of relative motion.

We also derive a relationship between the period at which the moving source emits successive wave crests ($dt_e'$) and the period at which a stationary observer receives them ($dt_r$) defining a Doppler factor $D = \dfrac{dt_r}{dt_e'}$. As long as the source approaches ($t_r < 0$) we have $D < 1$ advocating for a blue shift. If the source is reseeding ($t_r > 0$) we have $D > 1$ and a red shift takes place. Velocity and altitude influence the Doppler factor in a characteristic way. For $t_r = 0$ ($\theta = 90^0$) observer R detects the transversal Doppler shift. For very high values of $t_r$ ($[t_r] \to \infty$) we can consider that a longitudinal Doppler Effect takes place ($\theta \to 0^0$ and $\theta \to 180^0$ respectively).